\documentstyle[epsf,12pt]{article}
\newcommand{\AmS}{{\protect\the\textfont2
  A\kern-.1667em\lower.5ex\hbox{M}\kern-.125emS}}

\newcommand{\gsim}{{\protect
  \kern.18em\lower.5ex\hbox{$\stackrel{>}{\sim}$}\kern.25em}}
\newcommand{\lsim}{{\protect
  \kern.17em\lower.5ex\hbox{$\stackrel{<}{\sim}$}\kern.23em}}
\newcommand{\nn}{\nonumber}

\newcommand\be{\begin{equation}}
\newcommand\ee{\end{equation}}
\newcommand\bea{\begin{eqnarray}}
\newcommand\eea{\end{eqnarray}}

\newcommand{\phd}{\phi^{\dagger}}

\title{
{\vspace{-0cm} \normalsize
\hfill \parbox{40mm}{CERN 97-243}}\\[25mm]
Experiences with the\\
Polynomial Hybrid Monte Carlo Algorithm
       \thanks{
               Talk given by R.F. at the International
               Symposium on Lattice Field Theory, 22$-$26 July 1997,
               Edinburgh, Scotland}
        }

\author{
        Roberto Frezzotti \\ DESY \\ Notkestr. 85 \\
        D-22603 Hamburg, Germany                                  
\vspace{0.5cm} \\
        Karl Jansen \\ CERN                                  
        1211 Gen\`eve 23 \\ Switzerland \\
    }
\date{~}

\begin{document}

\maketitle

\begin{abstract}
We discuss a simulation algorithm for dynamical fermions, which 
combines the multiboson technique with the Hybrid Monte Carlo
algorithm. The algorithm turns out to give a substantial gain over
standard methods in practical simulations and to be suitable
for dealing with fermion zero modes in a
clean and controllable way.
\end{abstract}

\vspace{-1cm}

\section{Introduction}

In this contribution, we discuss a simulation algorithm, which we
call Polynomial Hybrid Monte Carlo (PHMC) \cite{PHMC_first}, for full lattice QCD,
implemented in the case of Wilson fermions. The algorithm is based
on two key ingredients, the interplay of which appears to be crucial
for a performance gain over the standard HMC algorithm.
The generation of gauge-field configurations is performed 
as first proposed in 
\cite{deF_Taka}: the usual inverse of the Dirac operator
is approximated (as in L\"uscher's multiboson algorithm 
\cite{multibos1})
by a polynomial in the operator and, with this polynomial defining the 
fermion action,
the simulation is done using a standard small-step Monte Carlo
method. 
The second crucial ingredient is the way of correcting for the above-mentioned
polynomial approximation. 
%(see \cite{Borr_deF} for alternative methods)
We suggest to do this 
by means of an efficient reweighing technique, which is reminescent
of earlier ideas, such as \cite{Bunk_et_al}, and allows us to deal in a clean and
controllable way with fermion zero modes, as discussed below.

\section{The PHMC algorithm}

Denoting by $U$ the configuration of lattice gauge links, the
expectation value of any gauge invariant observable ${\cal O}=
{\cal O}[U]$, in full QCD with $n_f =2$ degenerate flavours,
may be written as
\be
\langle {\cal O} \rangle = {\cal Z}^{-1} 
                               \left[ \int {\cal D}U 
e^{- S_g[U]} \mbox{det}(Q^2[U]) {\cal O}[U] \right]\; ,
\ee
where $S_g$ is the standard plaquette action for the pure gauge
sector ($\beta = 6/g_0^2$) and $Q$ is the Dirac operator for Wilson 
fermions multiplied by $\gamma_5$:
\bea
Q[U]_{x,y}=
c_0\gamma_5
\left[ \left(1+\frac{i}{2} \kappa c_{SW} \sigma_{\mu\nu}\hat{F}_{\mu\nu}\right)\right. 
\delta_{x,y}  \nn   \\
\left. - \kappa \sum_{\mu}P_{\mu}^{-}U_{x,\mu}\delta_{x+\mu,y}
+P_{\mu}^{+}U^{\dagger}_{x-\mu,\mu}\delta_{x-\mu,y}\right]
\eea
with $P_{\mu}^{\pm} = 1 \pm \gamma_{\mu}$, $\hat{F}_{\mu\nu}$ 
the standard `clover' discretization of $F_{\mu\nu}$, and $Q^\dagger = Q$.
%The term proportional to $c_{SW}$ is needed for the on-shell O($a$)
%improvement \cite{Improv} of the fermion action: $\hat{F}_{\mu\nu}$
%denotes the standard `clover' discretization of the field-strenght
%tensor. 
Here we present simulation results
only for $c_{SW}=0$. For $c_{SW}>0$ the computational cost 
per single molecular dynamics trajectory has been evaluated  and
performance tests are in progress.

We use the polynomial approximation of $(Q^2)^{-1}$ described in
\cite{Bunk_et_al}: the polynomial of degree $n$, denoted as
$P_{n,\epsilon}(s)$, approximates $s^{-1}$ in the range $\epsilon
< s <1$ with a relative fit error bounded by $\delta =
2 \left(\frac{1-\sqrt{\epsilon}}{1+\sqrt{\epsilon}}\right)^{n+1}$ 
(where $\epsilon > 0$). Choosing $c_0$ such that $\| Q^2\| <1$, we
may write the corresponding polynomial of $Q^2$ in a factorized
form:
\be
P_{n,\epsilon}(Q^2) = 
% C_{n,\epsilon} \prod_{k=1}^n (Q^2-z_k) =
C_{n,\epsilon} \prod_{k=1}^n (Q-r_k^*)(Q-r_k)\; , \label{fact_pol}
\label{fact_form}
\ee
where $ C_{n,\epsilon} $ is a positive constant, $r_k \equiv \sqrt{z_k} =
\mu_k + i\nu_k$ and the $z_k$'s are the complex roots of $P_{n,\epsilon}(s)$,
as in \cite{Bunk_et_al}. For the values of $n$ and $\epsilon$ used in our tests,
a careful ordering of the monomial factors appearing in (\ref{fact_pol}) was 
essential, at least on computers with 32-bit precision, in order to keep 
to a negligible level the 
rounding error in constructing $P_{n,\epsilon}(Q^2)$ using the 
factorized form, eq.(\ref{fact_pol}).                         
%\cite{Ordering}

%The fermion determinant ${\rm det}(Q^2)$ may be rewritten as
%$[{\rm det}(Q^2 P_{n,\epsilon}(Q^2))] [{\rm det}(P_{n,\epsilon}(Q^2))]^{-1}$
%and consequently 
The full QCD ($n_f=2$) partition function may
be represented as 
\bea 
{\cal Z} &
 = & \int {\cal D}U{\cal D}\phd{\cal D}\phi 
{\cal D}\eta^\dagger{\cal D}\eta
  W e^{- (S_g + S_P) } \nn \\
  S_P & = & S_P[U,\phi,\eta]=\phd P_{n,\epsilon}(Q^2[U]) \phi
     +\eta^\dagger\eta  \label{S_g_P}
\eea
by introducing the auxiliary pseudofermion fields (i.e. boson fields with
spin and colour indices) $\phi$, $\eta$ and the `correction'
factor $W=W[\eta,U]$:
\be
W= \exp\left\{\eta^\dagger (1-[Q^2\cdot P_{n,\epsilon}(Q^2)]^{-1}) \eta \right\} \; .
\ee

In the PHMC algorithm, the update of the gauge field configuration is
performed by using the $\Phi$-version \cite{Phi_alg} of the HMC algorithm
for the `approximate' but still non-local action $S_g + S_P$ of 
eq.~(\ref{S_g_P}). We denote averages in the theory with action $S_g + S_P$
as $\langle \dots \rangle_P$: reweighing with $W$ yields the true
averages, denoted as $\langle \dots \rangle$, for any observable ${\cal O} =
{\cal O}[U]$:
\be
\langle {\cal O} \rangle = \langle W \rangle_P^{-1}
\langle {\cal O}W \rangle_P  \label{true_ave} \; .
\ee    
The number ($N_{corr}$) of evaluations of $W$ per single molecular dynamics
trajectory (updating $U$) is relevant for the level of statistical error on
$\langle \cal O \rangle$, although $\langle \cal O \rangle$ itself is
correct, within statistical uncertainties,
for any value of $N_{corr}$. Each evaluation of $W$ requires a trivial
Gaussian update of the $\eta$-field and the solution of 
the system $[Q^2 P_{n,\epsilon}(Q^2)]\chi = \eta$.

\section{The results}

All the results discussed here, for both HMC and PHMC algorithms, have been obtained
using the even-odd preconditioned form 
% \cite{EO_precond} 
of the $Q$ operator, denoted 
by $\hat{Q}$, and a Sexton-Weingarten leap-frog integration scheme.
% \cite{SexWei}.
We adopted Schr\"odinger functional boundary conditions and monitored
few observables: the plaquette ($P$), the lowest ($\lambda_{min}$) and the highest
($\lambda_{max}$) eigenvalues of $\hat{Q}^2$ (normalized in such a way that 
$\lambda_{max}\lsim 1$).

Several tests on the $4^4$ lattice with different values of $(n,\epsilon)$ showed that
the chosen polynomial approximation should not be too bad, in order to avoid that
reweighing observables with $W$ induce large statistical fluctuations in true
averages. The approximation should not be too good either, 
in order to keep as low as possible the computational
cost per single trajectory (of length $\simeq 1$). In practice,
a reasonable compromise is obtained by choosing $\epsilon \simeq 2 \langle \lambda_{min}
\rangle $ and $n$ such that $\delta \simeq 0.01$--$0.02$, which means $n$ scaling
as $\epsilon^{-1/2}$. This criterion (plus short runs monitoring the statistical
fluctuations of $W$) allows us to quickly choose reasonable values of $(n,\epsilon)$. 
% for testing
%PHMC on a $8^4$ lattice, with the same bare parameters used for comparing multiboson
%and HMC algorithm in ref.~\cite{KJ_plen96}.  

%%%%%%%%%%%%%%%%%%%%%%%%%%%% TABLE %%%%%%%%%%%%%%%%%%%%%%%%%%%%%%%%%%%%%%%%%%%%%%%%%%%%%
\begin{table*}[hbt]
% space before first and after last column: 1.5pc
% space between columns: 3.0pc (twice the above)
\setlength{\tabcolsep}{1.5pc}
% -----------------------------------------------------
% adapted from TeX book, p. 241
% -----------------------------------------------------
%\vspace{2mm}
\label{tab:table1}
\begin{tabular*}{\textwidth}{@{}l@{\extracolsep{\fill}}ccccc}
\hline
$L^4$ & Algorithm $\quad \quad$& $C_{Q\phi}$  
&  $\sigma(P)/\langle P \rangle $ &  $\sigma(\lambda_{min})/\langle \lambda_{min} \rangle $  \\
\hline \hline
 $4^4$ & PHMC $\quad \quad$&  540   &  0.00024  &  0.0064 \\
 $4^4$ & HMC  $\quad \quad$&  868   &  0.00020  &  0.0057 \\
 $8^4$ & PHMC $\quad \quad$& 3974   &  0.00027  &  0.037\\
 $8^4$ & HMC  $\quad \quad$& 7398   &  0.00021  &  0.039 \\
\hline
\end{tabular*}
\caption{We give the single
trajectory cost, $C_{Q\phi}$, in units of $Q\phi$ operations.
We compare 
the relative statistical errors for $P$ and $\lambda_{min}$, obtained from a jack-knife 
blocking analysis, with $18000$ and $2745$ trajectories for $4^4$ and 
$8^4$ lattices, respectively. Note that we estimate a $10$ to $20$\% 
uncertainty on the statistical errors.}
%\caption{ bla ,bla}

\end{table*}
%%%%%%%%%%%%%%%%%%%%%%%%%%%%%%%%%%%%%%%%%%%%%%%%%%%%%%%%%%%%%%%%%%%%%%%%%%%%%%%%%%%%%%%%%%%%

In Table 1, we compare HMC and PHMC algorithms, as far as the computational cost per
single trajectory  and the relative statistical
error for $P$ and $\lambda_{min}$ are concerned. Bare lattice parameters are $\beta=6.4$,
$\kappa=0.15$ on the $4^4$ lattice. On the $8^4$ lattice we choose 
$\beta=5.6$ and  $\kappa=0.1585 \simeq \kappa_c$, 
which 
are the same bare parameters used for comparing the multiboson technique 
and the HMC algorithm in ref.~\cite{KJ_plen96}.  
The chosen values of $(n,\epsilon,N_{corr})$ are $(12,0.036,1)$ 
for the $4^4$ lattice, 
corresponding to one of the best choices, 
and $(48,0.026,2)$ for the $8^4$ lattice. 
The values of $\epsilon$ reflect a factor larger than
$10$ for the condition number of $\hat{Q}^2$,
between the $8^4$ and $4^4$ lattices.
% (defined as $\langle \lambda_{max} \rangle/\langle \lambda_{min} \rangle $), 
%which is about $700$ for the $8^4$ lattice case. 
Further details concerning the results reported in Table 1, such as 
the proper definition of the computational costs, molecular dynamics
parameters and average values with statistical uncertainties,
may be found in ref.~\cite{PHMC_first}. The formulae for computational costs 
appearing there are also valid for the ${\rm O}(a)$-improved fermions ($c_{SW}>0$). 
%, but in units of
%clover--$Q*\phi_k$ operations (\cite{KJ_plen96} and references therein).

Looking mainly at the $8^4$ lattice case, which is the one with 
a relatively large condition number of about $700$,
our results for the PHMC algorithm 
may be summarized as follows. First, average values of measured observables
always agree with the 
corresponding ones from the HMC algorithm (within statistical
uncertainties); without reweighing with $W$, systematic deviations, increasing with
$1-\langle W \rangle_P $, are observed, as expected.  Second, for suitable choices of
$(n, \epsilon)$, the statistical errors of measured observables
are the same as for the HMC algorithm, at least 
within their
relative uncertainty, which we estimate to be about $10$--$20\%$. 
At the same time, 
the computational cost per single trajectory, $C_{Q\phi}$, is almost a factor of 2 lower
than for the HMC algorithm. This is a consequence of the fact that $C_{Q\phi}$ is basically
proportional to $\epsilon^{-1}$ for the PHMC and to $\lambda_{min}^{-1}$ for the HMC algorithm.
Last but not least, we find that PHMC and HMC algorithms sample configuration space in a 
somewhat different way: for instance, the low-lying end of the distributions 
of $\lambda_{min}$ look different (Fig. 1).

%%%%%%%%%%%%%%%%%%%%%%%%%%%%%FIGURE%%%%%%%%%%%%%%%%%%%%%%%%%%%%%%%%%%%
\vspace{-0mm}
\begin{figure}[htb] \label{figure2}
\centerline{ \epsfysize=12.8cm
             \epsfxsize=12.8cm
             \epsfbox{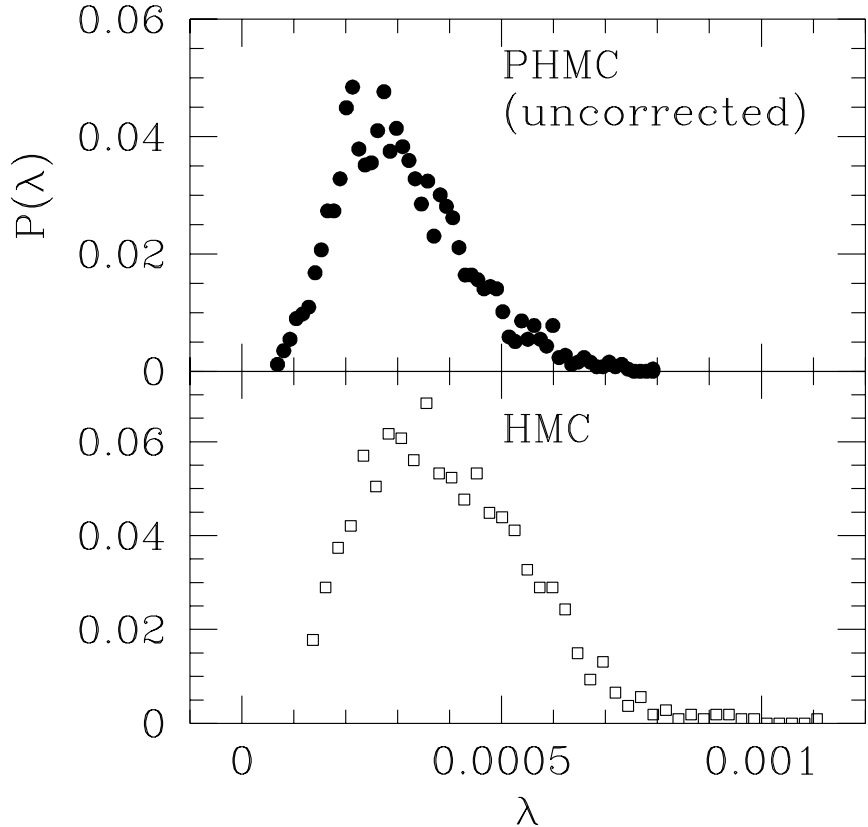}}
\vspace{-14mm}
\caption{The distribution of $\lambda = \lambda_{min}(\hat{Q}^2)$ as obtained
with equal statistics on the $8^4$ lattice
%from the generated set of $8^4$ gauge configurations ($\beta=5.6$ and $\kappa=0.1585$)
for both PHMC and HMC algorithms.
%PHMC(uncorrected) means that the PHMC distribution
%is not reweighed with the correction factor $W$.
}
\vspace{-0mm}
\end{figure}
%%%%%%%%%%%%%%%%%%%%%%%%%%%%%%%%%%%%%%%%%%%%%%%%%%%%%%%%%%%%%%%%%%%%%%

As expected from the properties of the polynomial approximation of $(\hat{Q}^2)^{-1}$, the
PHMC algorithm generates configurations with low-lying modes ($\lambda_{min} < \epsilon$)
of $(\hat{Q}^2)$ with a higher probability than algorithms like HMC or the multiboson 
technique made exact by an
accept/reject step \cite{Borr_deF}: in particular the PHMC algorithm generates, 
with small, non-zero
probability, configurations carrying fermion zero modes, which occur with vanishing
probability (i.e. never in a finite run time) when using the other algorithms
mentioned above. 
However, fermion zero modes give a finite, non-zero contribution to all those observables
where the divergence of quark propagators compensates for the vanishing of the probability
measure. In ref.~\cite{PHMC_first}, we discuss how
we may in principle deal with fermion zero modes, in order to evaluate the correction
factor $W$ and quark propagators, by making use of existing minimization algorithms (tested
up to $\lambda_{min} \sim 10^{-18}\lambda_{max}$). Therefore, the PHMC algorithm looks
particularly suitable to study the contribution of low-lying fermion
modes to physical observables in full QCD. For the same reasons, one may also
expect that the PHMC algorithm may overcome energy barriers, related to low-lying 
fermion modes, easier than the HMC algorithm,
leading to a better exploration of configuration space. 

This work is part of the ALPHA collaboration research programme.
We thank DESY for allocating computer time to this project.

\end{document}